\documentclass[%
 reprint,
 amsmath,amssymb,
 aps, prl
]{revtex4-1}

\usepackage{graphicx}
\graphicspath{}
\usepackage{float}
\usepackage{upgreek}
\usepackage{dcolumn}
\usepackage{bm}
\usepackage{hyperref}
\usepackage[mathlines]{lineno}
\usepackage{color}

 \newcommand{\be}{\begin{equation}}
 \newcommand{\ee}{\end{equation}}
 \newcommand{\bea}{\begin{eqnarray}}
 \newcommand{\eea}{\end{eqnarray}}

\begin{document}


\title{Universality of oscillating boiling in Leidenfrost transition}

\author{Mohammad Khavari$^{1,2}$}
\author{Tuan Tran$^1$}
\email{Corresponding author: ttran@ntu.edu.sg}
\affiliation{
$^1$ School of Mechanical \& Aerospace Engineering, 
	Nanyang Technological University, 
	50 Nanyang Avenue, 639798, Singapore \\
$^2$	Institute of Materials Research and Engineering, A*STAR, 3 Research Link, 117602, Singapore}
\date{\today}

\begin{abstract}
The Leidenfrost transition leads a boiling system 
to the boiling crisis, a state in which the liquid loses 
contact with the heated surface due to excessive vapor generation. 
Here, using experiments of liquid droplets boiling on 
a heated surface, 
we report  
a new phenomenon, termed oscillating boiling, at the Leidenfrost transition.
We show that oscillating boiling results from 
the competition between two effects: 
separation of liquid from the 
heated surface due to localized boiling,
and rewetting. 
We argue theoretically that the Leidenfrost transition can be predicted 
based on its link with the oscillating boiling phenomenon, 
and verify the prediction experimentally for various liquids.  
\end{abstract}

\maketitle
\bibliographystyle{apsrev4-1}

Boiling of liquid 
on a moderately heated surface 
removes heat effectively:
the liquid absorbs heat after touching the surface, vaporizes,
and the generated vapor is carried away by natural convection, 
letting liquid from the bulk replenish the surface. 
At elevated temperatures, this mechanism faces a 
fundamental problem, the so-called boiling crisis,
whereby 
excessive vapor completely eliminates 
liquid-surface contact (the Leidenfrost effect),
causing a severe drop in heat flux. 
Controlling occurrence of the Leidenfrost effect therefore 
is vital to either applications intolerant of the boiling crisis \cite{berthoud2000,*kim2007spray}, 
or those taking advantages of the liquid-surface separation \cite{linke2006self,*vakarelski2011drag,*vakarelski2012stabilization}. 
Nonetheless, 
despite its centuries-old history dating back to 1756 \cite{leiden1756}, 
the transition to the Leidenfrost regime remains
rich in empirical studies, 
but incomplete in physical understanding
\cite{bernardin1995validation,*bernardin2004leidenfrost,*mudawar2017}.

Research efforts aiming at 
understanding the Leidenfrost phenomenon 
have 
focused on 
the case of 
\emph{static} Leidenfrost droplets, i.e.,
droplets approaching 
a heated surface with negligible or small
initial velocity and subsequently floating on the surface.
The floating mechanism in this case
has been studied in great details: 
the viscous flow of vapor between 
a floating droplet and a heated surface
provides a counter force to 
the droplet's weight \cite{gottfried66,biance03,bernardin99,rein02}.
Similar hydrodynamical arguments have
also been used to explain the 
bouncing behavior without contact of droplets
falling on an unheated smooth surface with small velocity; 
the counter force in this case
is induced by the gas flow \cite{de2015wettability}. 
Although these theories
have been successfully used to 
explain the floating mechanism in the static case,
they have been assuming \emph{a priori} existence of the gas/vapor layer, 
thus precluding reference to the contact  
boiling behavior
and its role in the Leidenfrost transition. 
Their limitation is already hinted by 
experimental evidences
in the case of
\emph{dynamic} Leidenfrost droplets 
\cite{tran2012drop,khavari2015,shirota2016dynamic}, 
i.e., those approaching 
the heated surface with high velocity.
In such case, the Leidenfrost transition  
becomes independent of the impact velocity, 
signalling another separating mechanism 
instead of the one based on 
the gas/vapor flows. 
As a result, these theories cannot be used to 
reveal the mechanism of the Leidenfrost transition. 
The goal of this paper is to elucidate the Leidenfrost 
transition experimentally and theoretically.

\begin{figure}[t!]
\begin{center}
\includegraphics[width=0.47\textwidth]{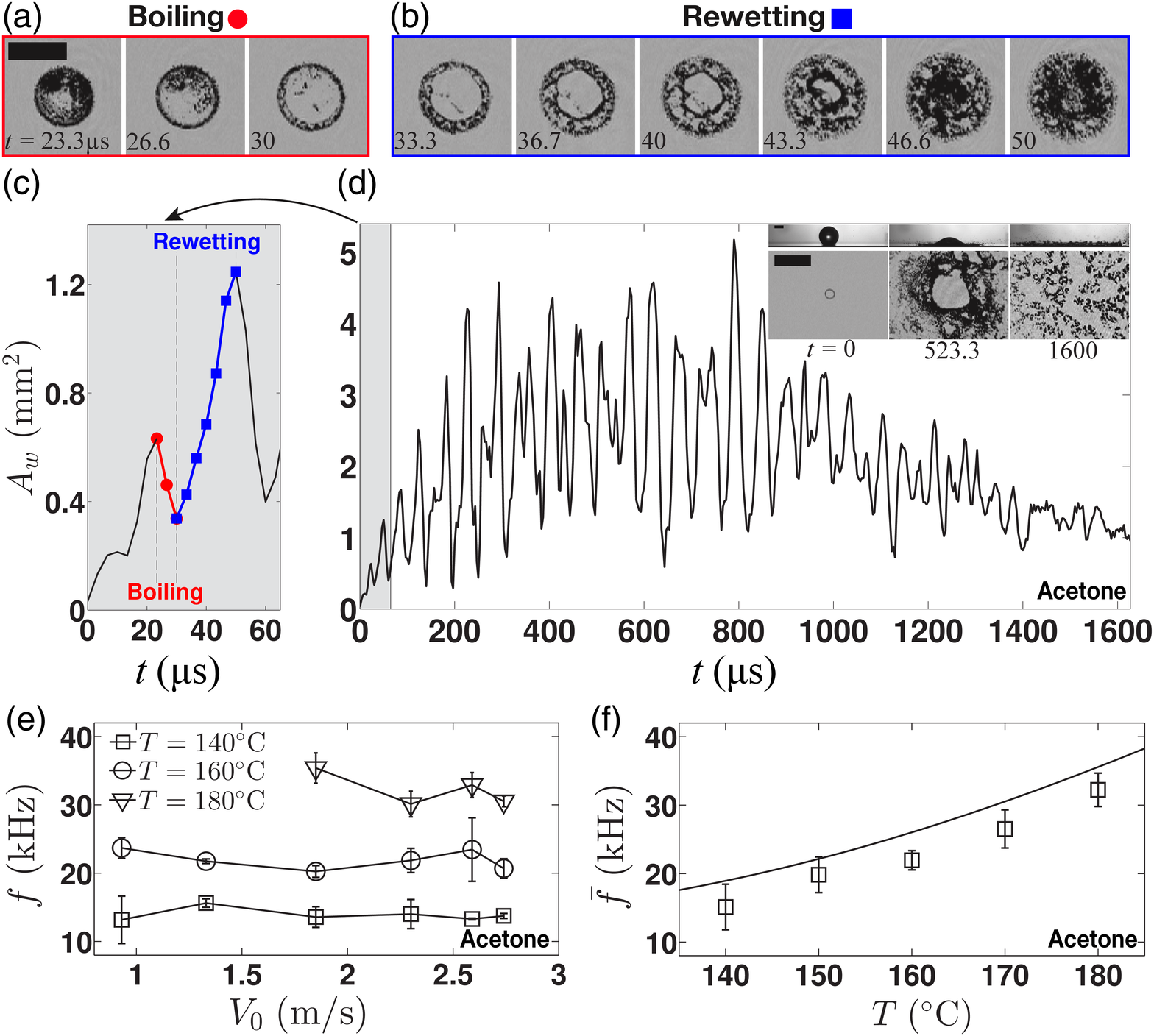}
\caption{\small{
Snapshots showing the wetted area,
measured by the total internal reflection (TIR) technique 
\cite{nagai1996,kolinski2012,khavari2015,shirota2016dynamic}, of
(a) the boiling process and 
(b) the rewetting process
for acetone droplets with velocity
 $V_0=2.7\,{\rm m \cdot s^{-1}}$ 
and surface temperature $T=150\,^\circ$C. 
(c) Wetted area $A_w$ vs. $t$ in the early stage of 
a droplet impacting on a heated sapphire surface. 
(d) Wetted area $A_w$ vs. $t$ during the entire impact process; 
the shaded area corresponds to the wetted area shown in (c).
Inset: snapshots show side-view (upper panel) 
and bottom-view (lower panel) of the impact at different times.  
(e) Frequency $f$ of oscillation vs. $V_0$ for 
different values of $T$ showing a weak dependence of $f$
on $V_0$. (f) Frequency $\bar{f}$ averaged across $V_0$
vs. $T$. The solid line is the frequency 
$V_{\rm re}/R_d$, where $V_{\rm re}$ is the rewetting velocity
(Eqn.~\ref{vre}), and $R_d\approx 1\,$mm the droplet radius.
All scale bars indicate the length scale of 1 mm.
  }} \label{fig1}
\end{center}
\vskip -0.9cm
\end{figure}

We show that the boiling behavior 
at the Leidenfrost transition is dominated 
by a new phenomenon, namely oscillating boiling. 
This boiling behavior results from the competition 
between two effects: 
separation of liquid from the heated 
surface due to localized boiling, and rewetting.

To show the oscillating boiling behavior, 
we visualize the surface's wetted areas and measure the total wetted area 
$A_w$ 
as a function of time $t$ (Fig.~\ref{fig1}). 
For droplet impacts in a wide range of velocity, liquids (see Table~\ref{tab1}), and 
surface temperatures close to the Leidenfrost transition, $A_w$ 
oscillates at remarkably high frequencies in the range
$12-32\,$kHz (Fig.~\ref{fig1}c and d). 
We observe that the frequency $f$ 
of oscillation depends weakly on the impact velocity $V_0$ 
(Fig.~\ref{fig1}e), but more significantly on the 
surface temperature $T$ (Fig.~\ref{fig1}f).


\begin{figure}[t!]
\begin{center}
\includegraphics[width=0.47\textwidth]{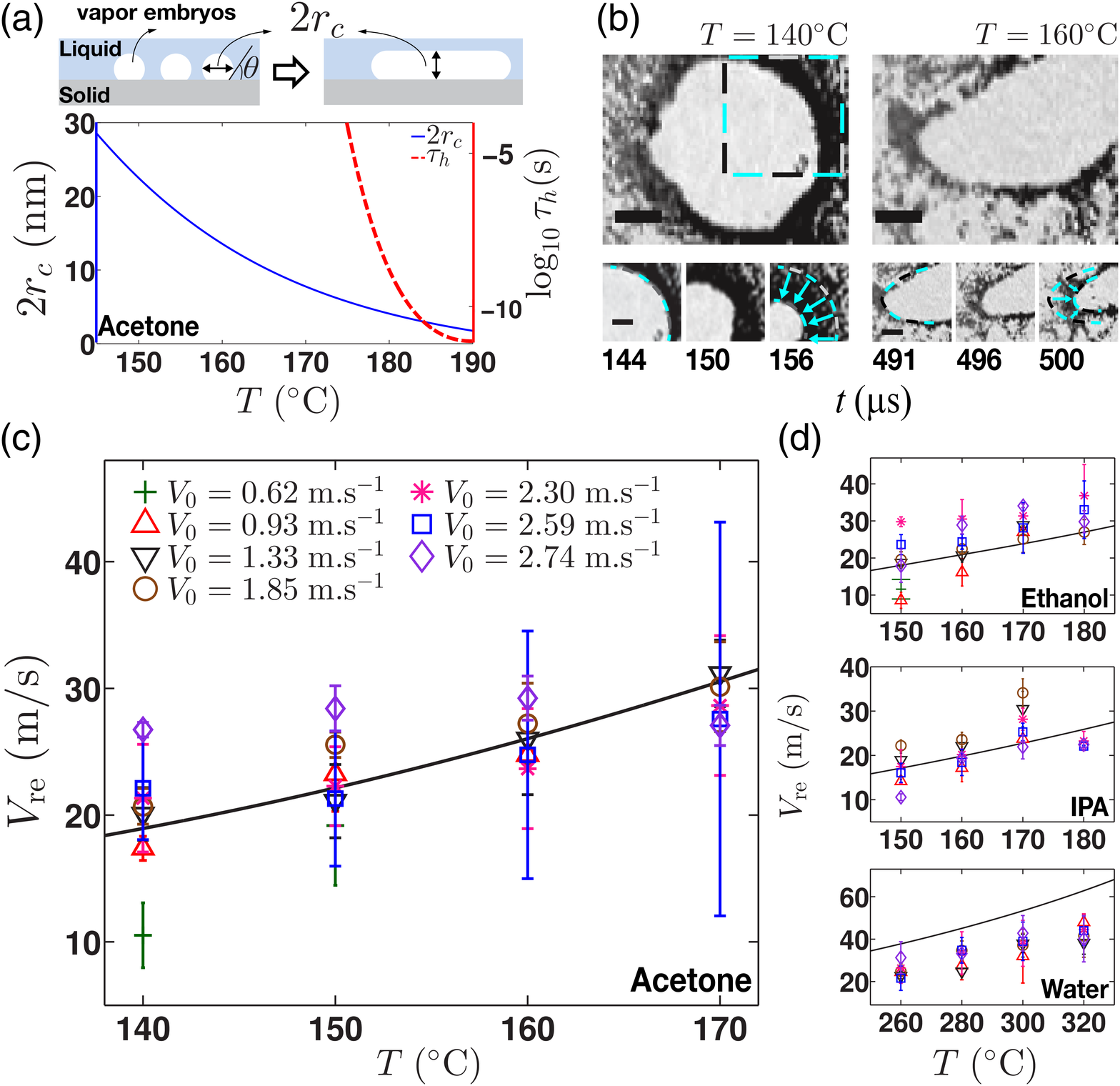}
\caption{\small{ 
(a) Critical size $2r_c$ of a vapor embryo (solid line) 
and evaporation time $\tau_h$ (dashed line) versus surface temperature $T$.
Upper panel: schematic showing vapor embryos
before and after merging to create a vapor layer. 
(b) Snapshots showing the rewetting process 
for acetone droplets impacting with velocity 
$V_0 = 2.3 \; \rm m \cdot s^{-1}$
at low temperature (left panel) and high temperature (right panel);
arrows indicate the motion of the three phase contact line. 
Scale bars represent the length scale of 0.3 mm.
The time stamps are measured from the first contact of impact. 
(c) Rewetting velocity $V_{\rm re}$ versus surface temperature
$T$ for acetone droplets for different impact velocities. 
(d)  $V_{\rm re}$ vs. $T$ for different impact velocities for 
ethanol, IPA, and water. 
Solid lines show the prediction according to Eqn.~\ref{vre}.
}}
\label{fig2}
\end{center}
\vskip -0.9cm
\end{figure}


The repeating pattern of the 
total wetted area $A_w$ typically consists of 
two distinct stages:
an abrupt drop and a 
subsequent increase in $A_w$ (Fig.~\ref{fig1}c). 
In the first stage,
exemplified in Fig.~\ref{fig1}a,
tiny dry spots
first appear 
spontaneously and rather uniformly
in the wetted area, then 
expand and merge 
to create larger dry areas. 
This observation suggests that the decrease in $A_w$ is
primarily caused by 
\emph{heterogeneous} boiling,
a process of forming the vapor 
from the liquid 
on a solid surface
at temperatures lower than 
boiling in liquid (\emph{homogeneous} boiling) \cite{carey92, cole1974boiling}. 
In the second stage, however, 
the wetted areas merge 
and invade the dry ones in the 
direction perpendicular to the three-phase contact line, 
signifying a rewetting process (Fig.~\ref{fig1}b). 
On this basis, 
we postulate that 
heterogeneous boiling and rewetting
are the two basic processes of the oscillating boiling behavior;
the rapid fluctuation of $A_w$ results 
from alternate domination 
of one process over the other (Movies S1-S6).
\begin{table}[]
	\centering
	\caption{Physical properties of the liquids at $20 ^\circ$C and atmospheric pressure \cite{faghri2006transport,lide2004CRC}, and experimental conditions.}
	\label{tab1}
		\begin{tabular}{clrrrr}

					& Unit 							& Acetone   	& Ethanol    	& IPA        	& Water    \\ 
		\hline
		$\sigma$ 		& $\rm mN \cdot m^{-1}$				& 23.7      		& 22.8		& 21.7	& 72.9      \\		
		$\rho_l$  		& $\rm kg \cdot m^{-3}$ 				& 790       		& 800                & 781	& 999       \\
		$T_b$ 		& $^\circ$C						& 56      		& 78			& 82		& 100       \\
		$h_{fg}$ 		& $\rm kJ \cdot kg^{-1}$				& 552       		& 1030		& 755	& 2454      \\
		$C_p$ 		& $\rm kJ \cdot kg^{-1}  \cdot K^{-1}$	& 2.16      		& 2.4			& 2.6		& 4.18      \\
		$\bar{M}$ 		& $\rm kg \cdot kmol^{-1}$			& 58.1      & 46                   & 60.1                 & 18        \\
		\hline
		$R_d$ 		& mm							& 1.0         	& 1.0                 & 1.0        & 1.2       \\
		$V_0$ 		& $\rm m \cdot s^{-1}$				& 0.6$-$2.7 	& 0.5$-$2.7	& 0.5$-$2.7	& 0.5$-$2.8 \\
		$T$ 			& $^\circ$C						& 20$-$370    	& 20$-$210	& 20$-$230	& 20$-$520    \\ 
		\hline
	\end{tabular}
\end{table}

\begin{figure*}[t!]
\begin{center}
\includegraphics[width=0.9\textwidth]{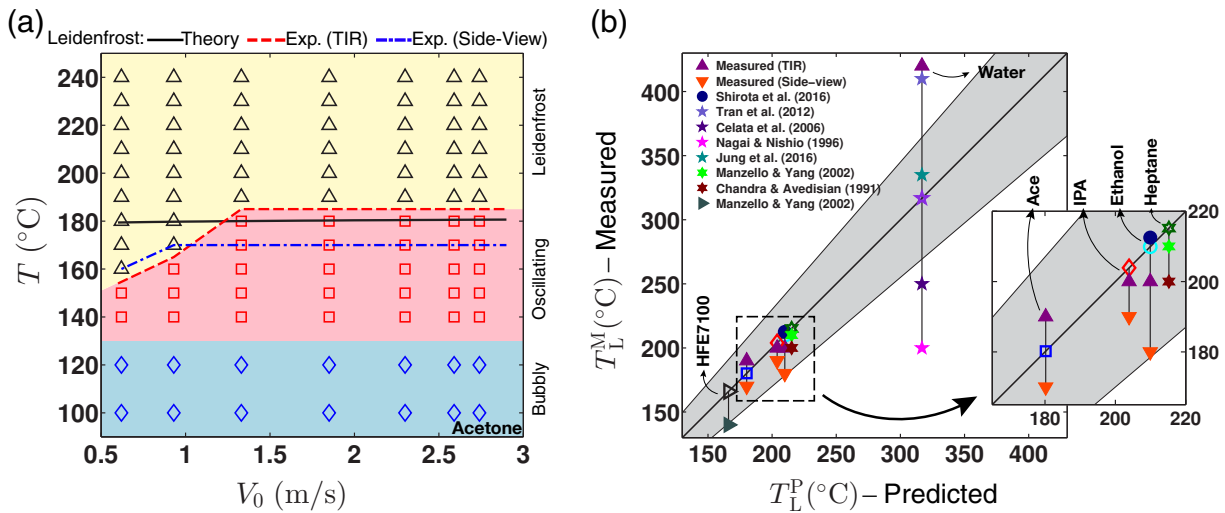}
\caption{\small{ 
(a) Phase diagram of characteristic boiling regimes
of acetone. 
The solid line represents the theoretical prediction of the Leidenfrost transition
(Eqn.~\ref{temp});
the dashed line and
the dashed-dotted line respectively mark the Leidenfrost transition
measured by TIR and side-view recordings. 
(b) Comparison between
the Leidenfrost temperatures 
obtained from theory ($T_{\rm L}^{\rm P}$) 
and those from experiments ($T_{\rm L}^{\rm M}$).
The open markers on the diagonal line 
indicate $T_{\rm L}^{\rm P}$. 
Upward solid triangles indicate $T_{\rm L}^{\rm M}$ obtained from TIR; 
downward solid triangles indicate $T_{\rm L}^{\rm M}$ from side-view recordings. 
All other solid markers are 
$T_{\rm L}^{\rm M}$ from previous studies \cite{shirota2016dynamic,tran2012drop, celata06,nagai1996,jung2016investigation,manzello2002experimental, *chandra91}.  
The shaded area indicates $\pm 15\%$ deviation from the theoretical values.
}}
\label{fig3}
\end{center}
\vskip -0.9cm
\end{figure*}

We now analyze the 
heterogeneous boiling process. 
On an ideally flat surface
immersed in a liquid at temperature $T_l$ and pressure $P_l$, 
vapor bubbles grow from embryos, tiny nanoscopic vapor pockets
in the form of spherical caps attaching to the surface (Fig.~\ref{fig2}a, schematics).
For an embryo to grow into a vapor bubble, 
the pressure difference across the vapor-liquid interface 
must overcome the Laplace pressure.   
Equivalently, its radius must be 
larger than a critical value
$r_c = 2\sigma F/(P_v - P_l)$, 
where $F$ is a correction factor to account for 
the partial spherical shape of the embryo \cite{note_f}.
Here, the vapor pressure $P_v$ inside the embryo 
depends on $T_l$ and $P_l$
as
$P_{v} = P_{sat}(T_l) \exp{\{[v_l(P_l - P_{sat}(T_l))]/RT_l}\}$ \cite{carey92},
where $v_l$ is the specific volume of the liquid, 
$P_{sat}(T_l)$ is the saturation pressure 
at $T_l$, and the liquid pressure $P_l$ 
is approximated using the atmospheric pressure. 
Owing to the very small size of embryos,
we assume that 
the temperature of liquid surrounding embryos is 
$T_l \approx T$, the surface temperature.
If we denote $J$ the generation rate per unit area of 
embryos having radius $r_c$,
$J$ can be readily calculated using the thermodynamic conditions 
of the liquid \cite{note_j,carey92}. 
It follows that the duration to populate a unit area 
by bubbles of radius $r_c$ is
$\tau_h = 1/J r_c^2$.
In Fig.~\ref{fig2}a, 
we show the plots of  
$2r_c$ and $\log_{10}(\tau_h)$ versus $T$
for acetone at the vicinity of Leidenfrost transition
($T_L = 180\,^\circ$C for acetone).
The reduction in 
$r_c$ indicates a decrease in thickness of
the vapor layer at higher $T$.
Remarkably, we also observe 
a sharp drop in $\tau_h$, implying that 
the required duration to generate a  
vapor layer becomes extremely small
as $T$ approaches the Leidenfrost transition. 
For instance, at $T = 180\,^\circ$C, $\tau_h = 1.6\,$ns.
This highlights the dominant role of 
the boiling process: a vapor layer is 
created immediately to separate the liquid 
from the surface if the liquid is sufficiently heated.  
However, we note that the boiling process takes away 
heat from the surface, causing its temperature to decrease,
and as a result, the heterogeneous boiling process 
is forced to stop.

As soon as the heterogeneous boiling process stops, 
the liquid in neighbouring wetted areas 
merges and rewets the dry areas
(Fig.~\ref{fig2}b). 
From the contact line's motion, indicated in Fig.~\ref{fig2}b,
we measure rewetting velocity 
$V_{\rm re}$ for different liquids and impact velocity $V_0$.
Generally, $V_{\rm re}$ is insensitive to $V_0$ and increases 
with surface temperature $T$ (see Fig.~\ref{fig2}c and d).
In order to understand this behavior, 
we note that a dry
area resulted from heterogeneous boiling 
is covered by a vapor layer of 
thickness $h \approx 2 r_c$ (Fig.~\ref{fig2}a, schematics).
Since the tested liquids have small viscosities,
the rewetting process is then 
driven by capillary pressure $\sigma/2r_c$
and resisted only by inertia $\rho_l V_{\rm re}^2$ \cite{wu04,winkels12}.
Here, both the surface tension $\sigma$ 
and density $\rho_l$ 
are functions of surface temperature $T$ \cite{note_para}.   
Balancing the capillary pressure and inertia
gives an estimate for the rewetting velocity
\be
V_{\rm re} \sim \left ( \frac{\sigma}{2\rho_l r_c} \right )^{1/2} = \left(\frac{P_v - P_l}{4\rho_l F} \right )^{1/2}. 
\label{vre}
\ee

In Fig.~\ref{fig2}c and d, we show several plots
of $V_{\rm re}$ versus $T$ for 
all tested liquids with impacting velocity $V_0$ 
varying from 0.6\,${\rm m\cdot s^{-1}}$ to 2.7\,${\rm m\cdot s^{-1}}$. 
We observe an excellent agreement with experimental results
for acetone, ethanol, IPA, and an overestimate for water.
We attribute possible drop in surface temperature
during impact to 
the discrepancy in the case of water:
due to its high latent heat and heat capacity compared to 
those of other liquids, 
the surface temperature in actuality 
may drop considerably during impact \cite{jung2016investigation}.
This causes an overestimation of the liquid temperature $T_l$,
as well as the vapor pressure $P_v(T_l)$, 
leading to an overestimation of $V_{\rm re}$ (Eqn.~\ref{vre}).
Nonetheless, these results not only suggest that 
the measured rewetting velocity
is consistent with 
our theoretical arguments for the rewetting process, but also
confirm that the vapor layer thickness is well described by 
the critical size of embryos. 
We conclude that the 
liquid-vapor dynamics 
in the oscillating boiling regime 
consist of two main processes: heterogeneous 
boiling followed by 
capillary-inertial rewetting. 

We are now ready to derive an expression to determine 
the \emph{dynamic} Leidenfrost transition, 
which marks the lower bound of the Leidenfrost regime
for impacting droplets; 
the surface temperature at this transition is termed as
\emph{dynamic} Leidenfrost temperature $T_L$.
We argue that at $T=T_L$,
the upward speed of vapor generation
must be comparable to the downward speed $V_0$ of liquid. 
Since the heterogeneous boiling process generates
a vapor layer of thickness $h=2r_c$ 
during the time $\tau_h$,
the upward speed of vapor generation is 
$V_v \sim 2r_c/\tau_h$.
Thus the dynamic Leidenfrost temperature $T_L$
satisfies the condition
\be
V_0 = \frac{2r_c(T_L)}{\tau_h(T_L)}.
\label{temp}
\ee
This condition
implicitly contains the dependence of $T_L$ on $V_0$ and the liquid properties, 
and we use it to determine the Leidenfrost transition
for our tested liquids.

We calculate $T_L$ for acetone using Eqn.~\ref{temp} 
and show a plot of $T_L$ vs. $V_0$ (solid line, Fig.~\ref{fig3})
together with a phase diagram 
consisting of three different characteristic behaviors:
Leidenfrost, 
\emph{oscillating} boiling, 
and bubbly boiling \cite{tran2012drop, khavari2015} (Fig.~\ref{fig3}).
We identify and categorize these behaviors
using TIR recordings 
of numerous impact experiments.
We note that although both the oscillating boiling and bubbly boiling 
behaviors allow solid-liquid contact, 
only the former exhibits high-frequency switching between the heterogeneous 
boiling and rewetting processes, while the boiling dynamics 
of the latter are much slower because of low surface temperature. 
In the phase diagram, 
the boundary separating the Leidenfrost and oscillating boiling regimes 
indicates the experimentally determined $T_L$ (dashed line),
which
increases monotonically with $V_0$ 
and reaches a plateau $T_L = 190\,^\circ$C
at $V_0=1.3\,{\rm m \cdot s^{-1}}$, 
consistent with earlier studies \cite{tran2012drop,shirota2016dynamic}.
We also show the Leidenfrost temperature determined 
by \emph{side-view} recordings (dashed-dotted line), 
a method incapable of directly detecting wetted areas during impact,
but instead relying on ejection of small droplets as
indication of wetted areas \cite{tran2012drop}.
The discrepancy between values of
$T_L$ determined by side-view recordings 
and ones by TIR technique, although not significant, is therefore 
expected. 
Nevertheless, the plateaued values of $T_L$ measured 
by both techniques are in remarkable agreement with
the predicted value $T_L = 180\,^\circ$C using Eqn.~\ref{temp},
highlighting that the proposed prediction for $T_L$ 
is valid for impacts at high velocity.

We emphasize that our prediction for $T_L$ 
is not applicable for the case of low impact velocity, 
e.g., $V_0\le1.3\,{\rm m\cdot s^{-1}}$,
where liquid-solid separation results from
the viscous stress induced by 
air/vapor flows under impacting droplets.
The viscous stress is
responsible for phenomena such as 
dimple formation \cite{tran2012drop,li2015}, 
bubble entrapment \cite{van2012direct,*bouwhuis2012maximal}, 
or even bouncing 
from unheated surfaces
for droplets at small impact velocity \cite{de2015wettability}.
We note that for $V_0\le 1.3\,{\rm m\cdot s^{-1}}$,
the compressible effect is negligible 
and the thickness of the air/vapor film under a
droplet decreases
with increasing $V_0$ \cite{li2015}.
The increase in $T_L$ with $V_0$ 
is expected  
to sustain the liquid-solid separation.
Thus the deviation of the predicted values of 
$T_L$ 
from the experimental ones 
at low impact velocity
reveals the region where 
the viscous stress caused by the air/vapor flows
becomes the dominant mechanism for separation. 

In Fig.~\ref{fig3}b, we show a plot comparing the
experimental Leidenfrost temperature ($T_L^M$) 
to the predicted one ($T_L^P$)
for six different liquids 
having broadly different thermal and 
physical properties.
For each liquid, we use the 
plateaued value of $T_L$ (shown in Fig.~\ref{fig3}b)
as the experimentally measured one.
The plot also consists 
of numerous experimental datasets from previous studies in which
measurements of the dynamic Leidenfrost temperature were reported;
most of them utilized the side-view technique to determine $T_L$.
Except for 
water, 
we observe that 
all the experimental values of $T_L$
are consistent, within $15\%$ deviation,
with the predicted ones;
typically the TIR measured values are
closer to theory than those obtained by the side-view technique.  
We note that 
while the surface temperature 
is assumed constant in our simplified theory,
it fluctuates considerably in reality
\cite{seki62,limbeek2016,shirota2016dynamic} as a result
of two competing heat transfer processes: 
one from the solid surface to the liquid, 
and the other from the solid bulk to the surface.
In addition, variations in surface roughness and 
impact velocity may contribute to discrepancy 
between $T_L^M$ and $T_L^P$.
In the case of water, which has exceptionally
high heat capacity and latent heat compared to other tested liquids (Table~\ref{tab1}), 
we expect a more severe 
drop in surface temperature,
which may be the main cause of the
vastly disparate reported values for $T_L^M$
(from $200\,^\circ {\rm C}$ to $410\,^\circ {\rm C}$).
The reduction in surface temperature, however, 
implies that the theory underestimates
$r_c$ and subsequently overestimates
$V_{\rm re}$, consistent with 
the result shown in Fig.~\ref{fig2}d for water.
In other words, the prediction of $T_L$ for water 
is missing a prefactor to account for the substantial 
drop in surface temperature.
A detailed analytical prediction of $T_L$ 
for water, therefore, 
must be 
derived on the ground of non-negligible fluctuation in surface temperature
and merits further investigation.


From our experimental measurements and
analysis of $T_L$,
we infer that there are two
fundamentally different mechanisms for 
Leidenfrost transition  
at two extremes of velocity.
At low impact velocity, 
the separation mechanism 
is mainly associated the
viscous stress induced in the air/vapor flows.
By contrast, at high impact velocity,
the mechanism for Leidenfrost transition
is dictated by the  
heterogeneous boiling process, 
i.e., the kinetic limit of superheat at 
the solid-liquid interface.
This explains the asymptotic behavior
of the Leidenfrost transition 
at high 
impact velocity.
The heterogeneous boiling process 
also causes the oscillating boiling behavior, i.e., 
high-frequency fluctuation of 
the wetted area due to alternative 
domination of either 
heterogeneous boiling or rewetting
over the other. 
The transition to 
Leidenfrost regime, therefore, 
reduces to the sustaining capability 
of the heterogeneous boiling process.
For practical boiling applications at high temperatures, 
a direct implication from our theory is that in order to avoid 
the boiling crisis, 
the properties of both the liquid and the boiling surface 
should be selected
to obtain high rewetting velocity and subsequently 
to sustain heterogeneous boiling. 
Furthermore, the 
role of capillary-inertial rewetting in the transition to boiling crisis on smooth surfaces
can be further extended to design rough surfaces 
used in boiling.
We conclude by highlighting that our findings offer a
theoretical framework 
to treat 
the Leidenfrost transition, a crucial step 
toward achieving complete control 
of the Leidenfrost effect.

\section*{ACKNOWLEDGMENTS}
We thank H. Kellay and P. Chakraborty for helpful discussions.
This work was 
 funded by Nanyang Technological University and A*STAR, Singapore.
 M.\ Khavari acknowledges SINGA scholarship from A*STAR.
\vskip -0.5cm



\begin{thebibliography}{39}
\bibitem [{not({\natexlab{a}})}]{note_f}%
  \BibitemOpen
  \href@noop {} {}\bibinfo {note} {The factor $F = 0.5 +
  0.75\cos{\theta} - 0.25\cos^3{\theta}$ is the correction factor that accounts
  for partially wettable liquids with contact angle $\theta$.}\BibitemShut
  {Stop}%
\bibitem [{not({\natexlab{b}})}]{note_j}%
  \BibitemOpen
  \href@noop {} {}\bibinfo {note} {The embryo formation
  rate $J$ relates to liquid properties and boiling conditions as
  \cite{carey92} $J = \Gamma_1\, \Gamma_2 \, \Gamma_3$, where $\Gamma_1 =
  \rho_{N,l}^{2/3} (1+\cos{\theta} )/2F$, $\Gamma_2 = [3F\sigma_{lv} / (\pi m)
  ]^{1/2}$, $\Gamma_3 = \exp \{-16\pi F \sigma_{lv}^3 / \{3 k_B T_l \, [\eta
  P_{sat}(T_l) - P_l]^2 \} \}$, and $ \eta = \exp\displaystyle\{v_l\, [P_l -
  P_{sat}(T_l)]/(R T_l)\}$. Here $\rho_{N,l} = N_A/(v_l \bar{M})$ is the number
  density of liquid molecules per unit volume, $m = \bar{M}/N_A$ the mass of
  one molecule, $\bar{M}$ the molecular mass in ${\rm kg \cdot {kmol}^{-1}}$,
  $N_A = 6.02\times 10^{26}\,{\rm kmol}^{-1}$ the Avogadro constant, $k_B =
  1.38\times 10^{-23}\,{\rm J\cdot K^{-1}}$ the Boltzmann constant, $R
  =\bar{R}/\bar{M}$ the specific gas constant in ${\rm J\cdot {kg}^{-1} \cdot
  K^{-1}}$, and $\bar{R} = 8314 .4\, {\rm J \cdot K^{-1}\cdot kmol^{-1}}$ the
  universal gas constant.}\BibitemShut {Stop}%
\end{thebibliography}

%

\end{document}